# Visualization of Very Large High-Dimensional Data Sets as Minimum Spanning Trees

Daniel Probst[1,*] and Jean-Louis Reymond[1,*]

[1]*Department of Chemistry and Biochemistry, National Center for Competence in Research NCCR TransCure, University of Berne, Freiestrasse 3, 3012 Berne, Switzerland*

[*]e-mail: daniel.probst@dcb.unibe.ch, jean-louis.reymond@dcb.unibe.ch

## Abstract

The chemical sciences are producing an unprecedented amount of large, high-dimensional data sets containing chemical structures and associated properties. However, there are currently no algorithms to visualize such data while preserving both global and local features with a sufficient level of detail to allow for human inspection and interpretation. Here, we propose a solution to this problem with a new data visualization method, TMAP, capable of representing data sets of up to millions of data points and arbitrary high dimensionality as a two-dimensional tree (http://tmap.gdb.tools). Visualizations based on TMAP are better suited than t-SNE or UMAP for the exploration and interpretation of large data sets due to their tree-like nature, increased local and global neighborhood and structure preservation, and the transparency of the methods the algorithm is based on. We apply TMAP to the most used chemistry data sets including databases of molecules such as ChEMBL, FDB17, the Natural Products Atlas, DSSTox, as well as to the MoleculeNet benchmark collection of data sets. We also show its broad applicability with further examples from biology, particle physics, and literature.

## Introduction

The recent development of new and often very accessible frameworks and powerful hardware has enabled the implementation of computational methods to generate and collect large high dimensional data sets and created an ever increasing need to explore as well as understand these data.[1–9] Generally, large high-dimensional data sets are matrices where rows are samples and columns are measured variables, each column defining a dimension of the space which contains the data. Visualizing such data sets is challenging because reducing the dimensionality, which is required in order to make the data visually interpretable for humans, is both lossy and computationally expensive.[10]

Large high-dimensional data sets are frequently used in the chemical sciences. For instance the ChEMBL database ($n = 1{,}159{,}881$) of bioactive molecules from the scientific literature and their associated biological assay data are used daily in the area of drug discovery.[11] Further examples of large databases containing molecules include FDB17 ($n = 10{,}101{,}204$), a fragment-like subset of the enumerated database GDB17 listing theoretically possible molecules up to 17 atoms,[12–14] and DSSTox ($n = 848{,}816$), containing molecules investigated for toxicity.[15] Examples of smaller data sets include the Natural Products Atlas ($n = 24{,}594$), collecting microbially-derived natural products;[16] Drugbank ($n = 9{,}300$), listing molecules marketed or investigated as drugs;[17] and the MoleculeNet benchmark, containing a collection of 16 data sets of small organic molecules.[18]

To visualize such databases, simple linear dimensionality reduction methods such as principal component analysis and similarity mapping readily produce 2D- or 3D-representations of global features.[19–25] However, local features defining the relationships between close or even nearest neighbor (NN) molecules, which are very important to understand the structure of data, are mostly lost, limiting the applicability of linear dimensionality reduction methods for visualization. The important NN relationships are much better preserved using non-linear manifold learning algorithms, which assume that the data lies on a lower-dimensional manifold embedded within the high-dimensional space. Algorithms such as nonlinear principal component analysis (NLPCA or autoencoders), t-distributed stochastic neighbor embedding (t-SNE), and more recently uniform manifold approximation and projection (UMAP) are based on this assumption.[26–28] Other techniques used are probabilistic generative topographic maps (GTM) and self-organizing maps (SOM), which are based on artificial neural networks.[29,30] However, these algorithms have time complexities between at least $O(n^{1.14})$ and $O(n^5)$, limiting the size of to be visualized data sets.[31] The same limitations in terms of data set size apply when distributing data in a tree by implementing the neighbor joining algorithm or similar methods used to create phylogenetic trees.[32,33] This limiting behavior has been documented by the ChemTreeMap tool, which can only visualize up to approximately 10,000 data points (molecules or clusters of molecules).[34] Due to the described challenges, large scientific data sets are generally visualized in aggregated or reduced form.[35,36]

Here we present an algorithm, named TMAP (Tree MAP), to generate and distribute intuitive visualizations of large data sets in the order of up to $10^7$ with arbitrary dimensionality in a tree. Our method is based on a combination of locality sensitive hashing, graph theory, and modern web technology which also integrates into established data analysis and plotting workflows. This tree-based layout facilitates visual inspection of the data with a high resolution by explicitly visualizing the closest distance between clusters and the detailed structure of clusters through branches and sub-branches. We demonstrate the performance of TMAP with toy data sets from computer graphics and with ChEMBL subsets of different size and composition, and show that it surpasses comparable algorithms such as t-SNE and UMAP in terms of time and space complexity. We further exemplify the use of TMAP for visualizing large high-dimensional data sets from chemistry as well as from further scientific fields (Table 1).

**Table 1** Data sets visualized using TMAP.

| Data Set | Description | Data Type | Size |
|---|---|---|---|
| **Toy Data Sets** | | | |
| COIL20 | Gray-scale images of 20 objects, each rotated 72x at 5° intervals. | Images | 1,440 |
| MNIST | Gray-scale images of handwritten digits. | Images | 70,000 |
| Fashion MNIST | Gray-scale images of fashion items from 10 classes. | Images | 70,000 |
| **Chemical Compound Databases and PDB** | | | |
| ChEMBL | Bioactive molecules with drug-like properties. | SMILES | 1,159,881 |
| FDB17 & ChEMBL | Fragment database (up to 17 atoms) and ChEMBL. | SMILES | 11,261,085 |
| Natural Products Atlas | Bacterial and fungal natural products. | SMILES | 24,594 |
| DSSTox | U.S. EPA information on toxicity of chemicals. | SMILES | 848,816 |
| PDB | Information on the 3D structures of proteins and nucleic acids. | Atomic coordinates | 131,236 |
| Drugbank | Approved, investigational, experimental, and withdrawn drugs. | SMILES | 9,300 |
| **MoleculeNet Benchmark Data Sets** | | | |
| QM8 | Subset of GDB-13 with associated QM properties. | SMILES | 21,786 |
| QM9 | Subset of GDB-13 with associated QM properties. | SMILES | 133,885 |
| ESOL | Common organic small molecules with solubility information. | SMILES | 1,128 |
| FreeSolv | Calculated and experimental hydration free energy of molecules. | SMILES | 642 |
| Lipophilicity | Experimental results of logD for organic small molecules. | SMILES | 4,200 |
| PCBA | PubChem subset with biological activities. | SMILES | 437,929 |
| MUV | PubChem subset for virtual screening validation. | SMILES | 93,087 |
| HIV | Experimental results for HIV replication inhibition. | SMILES | 41,127 |
| PDBind | Binding affinities for ligands in biomolecular complexes. | SMILES | 11,908 |
| BACE | IC50 values against BACE-1 (human β-secretase 1). | SMILES | 1,513 |
| BBBP | Ability of organic molecules to cross the blood-brain barrier. | SMILES | 2,039 |
| Tox21 | Toxicity measurements on 12 targets. | SMILES | 7,831 |
| ToxCast | Toxicity measurements on more than 600 targets. | SMILES | 8,575 |
| SIDER | Adverse drug reactions of a selection of marketed drugs. | SMILES | 1,427 |
| ClinTox | FDA approved drugs that failed clinical trials for toxicity reasons. | SMILES | 1,478 |
| **Other Data Sets** | | | |
| PubMed Central | Full-text archive of biomedical and life sciences journal literature. | Text | 327,628 |
| Gutenberg | A subset of public domain Project Gutenberg eBooks. | Text | 3,036 |
| NIPS | Abstracts of NIPS conference papers from 1987-2015. | Text | 7,241 |
| RNA Sequencing | A subset of the PANCAN database. | Gene expression | 801 |
| ProteomeHD | Human proteome co-regulation data. | Co-regulation scores | 5,013 |
| Flowcytometry | Data gathered from a flow cytometry experiment. | Signal intensity | 436,877 |
| MiniBooNE | Data gathered by the MiniBooNE particle physics experiment. | Particle ID | 130,065 |

.

## Methods

Given an arbitrary data set as an input, TMAP encompasses four phases: (I) LSH forest indexing,[37,38] (II) construction of a $c$-approximate $k$-nearest neighbor graph, (III) calculation of a minimum spanning tree (MST) of the $c$-approximate $k$-nearest neighbor graph,[39] and (IV) generation of a layout for the resulting MST.[40]

During phase I, the input data are indexed in an LSH forest data structure, enabling $c$-approximate $k$-nearest neighbor ($k$-NN) searches with a time complexity sub-linear in $n$. Text and binary data are encoded using the MinHash algorithm, while integer and floating-point data are encoded using a weighted variation of the algorithm.[41–43] The LSH Forest data structure for both MinHash and weighted MinHash data is initialized with the number of hash functions $d$ used in encoding the data, and the number of prefix trees $l$. An increase in the values of both parameters led to an increase in main memory usage; however, higher values for $l$ also decrease query speed. The effect of parameters $d$ and $l$ on the final visualization is shown in Fig. S1. The use of a combination of (weighted) MinHash and LSH Forest, which supports fast estimation of the Jaccard distance between two binary sets, has been shown to perform very well for molecules.[44] Note that other data structures and algorithms implementing a variety of different distance metrics may show better performance on other data and can be used as drop-in replacements of phase I.

In phase II, an undirected weighted $c$-approximate $k$-nearest neighbor graph ($c$-$k$-NNG) is constructed from the data points indexed in the LSH forest, where an augmented variant of the LSH forest query algorithm we previously introduced for virtual screening tasks is used to increase efficiency.[45] The $c$-$k$-NNG construction phase takes two arguments, namely $k$, the number of nearest-neighbors to be searched for, and $k_c$, the factor used by the augmented query algorithm. The variant of the query algorithm increases the time complexity of a single query from $O(\log n)$ to $O(k \cdot k_c + \log n)$, resulting in an overall time complexity of $O(n(k \cdot k_c + \log n))$, where practically $k \cdot k_c > \log n$, for the $c$-$k$-NNG construction. The edges of the $c$-$k$-NNG are assigned the Jaccard distance of their incident vertices as their weight. Depending on the distribution and the hashing of the data, the $c$-$k$-NNG can be disconnected (1) if outliers exist which have a Jaccard distance of 1.0 to all other data points and are therefore not connected to any other nodes or (2) if, due to highly connected clusters of size $\geq k$ in the Jaccard space, connected components are created. However, the following phases are agnostic to whether this phase yields a disconnected graph. The effect of parameters $k$ and $k_c$ on the final visualization is shown in Fig. S2. Alternatively, an arbitrary undirected graph can be supplied to the algorithm as a (weighted) edge list.

During phase III, a minimum spanning tree (MST) is constructed on the weighted $c$-$k$-NNG using Kruskal's algorithm, which represents the central and differentiating phase of the described algorithm. Whereas comparable algorithms such as UMAP or t-SNE attempt to embed pruned graphs, TMAP removes all cycles from the initial graph using the MST algorithm, significantly lowering the computational complexity of a low dimensional embedding. The algorithm reaches a globally optimal solution by applying a greedy approach of selecting locally optimal solutions at each stage—properties which are also desirable in data visualization. The time complexity of Kruskal's algorithm is $O(E + \log V)$, rendering this phase negligible compared to phase II in terms of execution time. In the case of a disconnected $c$-$k$-NNG, a minimum spanning forest is created.

Phase IV lays out the tree on the Euclidean plane. As the MST is unrooted and to keep the drawing compact, the tree is not visualized by applying a tree but a graph layout algorithm. In order to draw MSTs of considerable size (millions of vertices), a spring-electrical model layout algorithm with multilevel multipole-based force approximation is applied. This algorithm is provided by the open graph drawing framework (OGDF), a modular C++ library.[40] In addition, the use of the OGDF allows for effortless adjustments to the graph layout algorithm in terms of both aesthetics and computational time requirements. Whereas several parameters can be configured for the layout phase, only parameter $p$ must be adjusted based on the size of the input data set (Fig. S3). This phase constitutes the bottleneck regarding computational complexity.

## Results and Discussion

### TMAP performance assessment with Toy Data Sets and ChEMBL Subsets

The quality of our TMAP algorithm is first assessed by comparing TMAP and UMAP to visualize the common benchmarking data sets MNIST, FMNIST, and COIL20 (Fig. 1). UMAP generally represents clusters as tightly packed patches and tries to reach maximal separation between them. On the other hand, TMAP visualizes the relations between, as well as within, clusters as branches and sub-branches. While UMAP can represent the circular nature of the COIL20 subsets, TMAP cuts the circular clusters at the edge of largest difference and joins subsets through one or more edges of smallest difference (Fig. 1a, b). However, the plot shows that this removal of local connectivity leads to an untangling of highly similar data (shown in dark green, orange, dark red, dark purple, and light blue). This behavior has been assessed and compared to UMAP in Figures S4 and S5, where it is shown that both TMAP and UMAP have to sacrifice locality preservation for more complex examples. For the MNIST and FMNIST data sets, the tree structure results in a higher resolution of both variances and errors within clusters as it becomes apparent how sub clusters (branches within clusters) are linked and which true positives connect to false positives (Fig 1c, d, e, f).

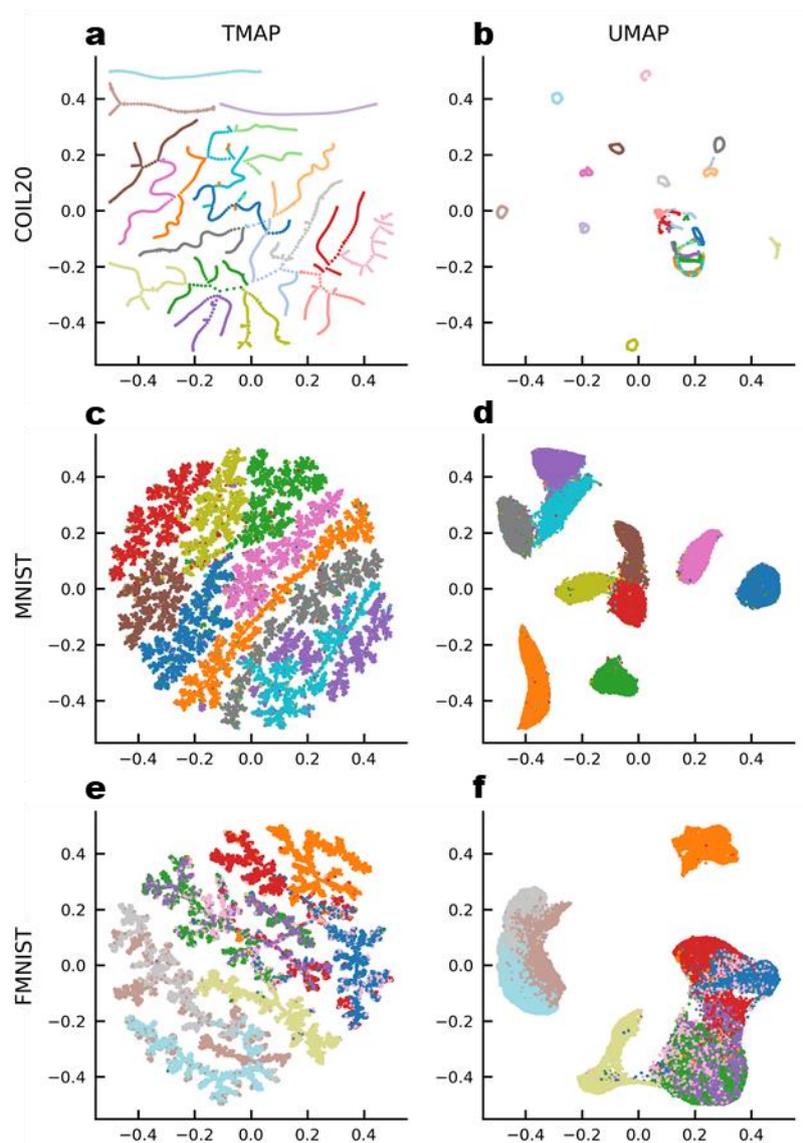

**Fig. 1. Comparison between TMAP and UMAP on benchmark data sets**. Please use the interactive versions of the TMAP visualizations at http://tmap.gdb.tools to see images associated with each point on the map. TMAP explicitly visualizes the relations between as well as within clusters. (**a**, **b**) While UMAP represents the circular nature of the COIL20 subsets, TMAP cuts the circular clusters at the edge of largest difference and joins clusters through an edge of smallest difference. (**c**, **d**, **e**, **f**) For the MNIST and FMNIST data sets, the tree structure allows for a higher resolution of both variances and errors within clusters as it becomes apparent how sub clusters (branches within clusters) are linked and which true positives connect to false positives. The image data of all three sets was binarized using the average intensity per image as a threshold.

In a second, more applied comparison example, we visualize data from ChEMBL using TMAP and UMAP. For this analysis molecular structures are encoded using ECFP4 (extended connectivity fingerprint up to 4 bonds, 512-D binary vector), a molecular fingerprint encoding circular substructures and which performs well in virtual screening and target prediction.[46–48] We consider a subset $S_t$ of the top 10,000 ChEMBL compounds by insertion date, as well as a random subset $S_r$ of 10,000 ChEMBL molecules.

Taking the more homogeneous set $S_t$ as an input, the 2D-maps produced by each representation, plotted using the Python library matplotlib, illustrate that TMAP, which distributes clusters in branches and subbranches of the MST, produces a much more even distribution of compounds on the canvas compared to UMAP, thus enabling better visual resolution (Fig. 2a, b). Furthermore, in a visualization of the heterogeneous set $S_r$, nearest neighbor relationships (locality) are better preserved in TMAP compared to UMAP, as illustrated by the positioning of the 20 structurally nearest neighbors of compound CHEMBL3701602,[49] reported as a potent inhibitor of human tyrosine-protein kinase SYK. The 20 structurally similar nearest neighbors are defined as 20 nearest neighbors in the original 512-dimensional fingerprint space. TMAP directly connects the query compound to three of the 20 nearest neighbors, CHEMBL3701630, CHEMBL3701611, and CHEMBL38911457, its nearest, second nearest, and 15[th] nearest neighbor respectively. The nearest neighbors 1 through 7 are all within a topological distance of 3 around the query (Fig. 2c). In contrast, UMAP has positioned nearest neighbors 2, 3, 9, and 18, among several even more distant data points, closer to the query than the nearest neighbor from the original space (Fig. 2d). Indeed, TMAP preserves locality in terms of retaining 1-nearest neighbor relationships much better than UMAP, applying both topological and Euclidean metrics (Fig. 2e, f; Fig. S6). The quality of the preservation of locality largely depends on parameter $d$, with adjustments to parameters $k$ and $k_c$ only having a minor influence (Fig. S7). Moreover,

TMAP yields reproducible results when running on identical parameters and input data, whereas results of comparable algorithms such as UMAP change considerably with every run (Fig. S8).[26]

In terms of calculation times, TMAP and UMAP have comparable running time $t$ and memory usage $a$ for small random subsets of the 512-D ECFP-encoded ChEMBL data set with sizes $n = 10,000$ and $n = 100,000$, TMAP significantly outperforms UMAP for larger random subsets ($n = 500,000$ and $n = 1,000,000$) (Fig. 2h, i). Further insight into the computational behavior of TMAP is provided by analyzing running times for the different phases based on a larger subset ($n = 1,000,000$) of the ECFP4-encoded ChEMBL data set (Fig. 2g). During phase I of the algorithm, which accounts for $180s$ of the execution time and approximately $5GB$ of main memory usage, data is loaded and indexed in the LSH Forest data structure in chunks of $100,000$, as expressed by $10$ distinct jumps in memory consumption. The construction of the $c$-$k$-NNG during phase II requires a negligible amount of main memory and takes approximately $110s$. During $10$ seconds of execution time, MST creation (phase III) occupies a further $2GB$ of main memory of which approximately $1GB$ is retained to store the tree data structure. The graph layout algorithm (phase IV) requires $2GB$ throughout $55s$, after which the algorithm completes with a total wall clock run time of $355s$ and peak main memory usage of $8.553GB$.

Note that TMAP supports Jaccard similarity estimation through MinHash and weighted MinHash for binary and weighted sets, respectively. While the Jaccard metric is very suitable for chemical similarity calculations based on molecular fingerprints, the metric may not be the best option available to problems presented by other data sets. However, there exists a wide range of LSH families supporting distance and similarity metrics such as Hamming distance, $l_p$ distance, Levenshtein distance, or cosine similarity, which are compatible with TMAP.[50,51] Furthermore, the modularity of TMAP allows to plug in arbitrary nearest-neighbor-graph creation techniques or load existing graphs from files.

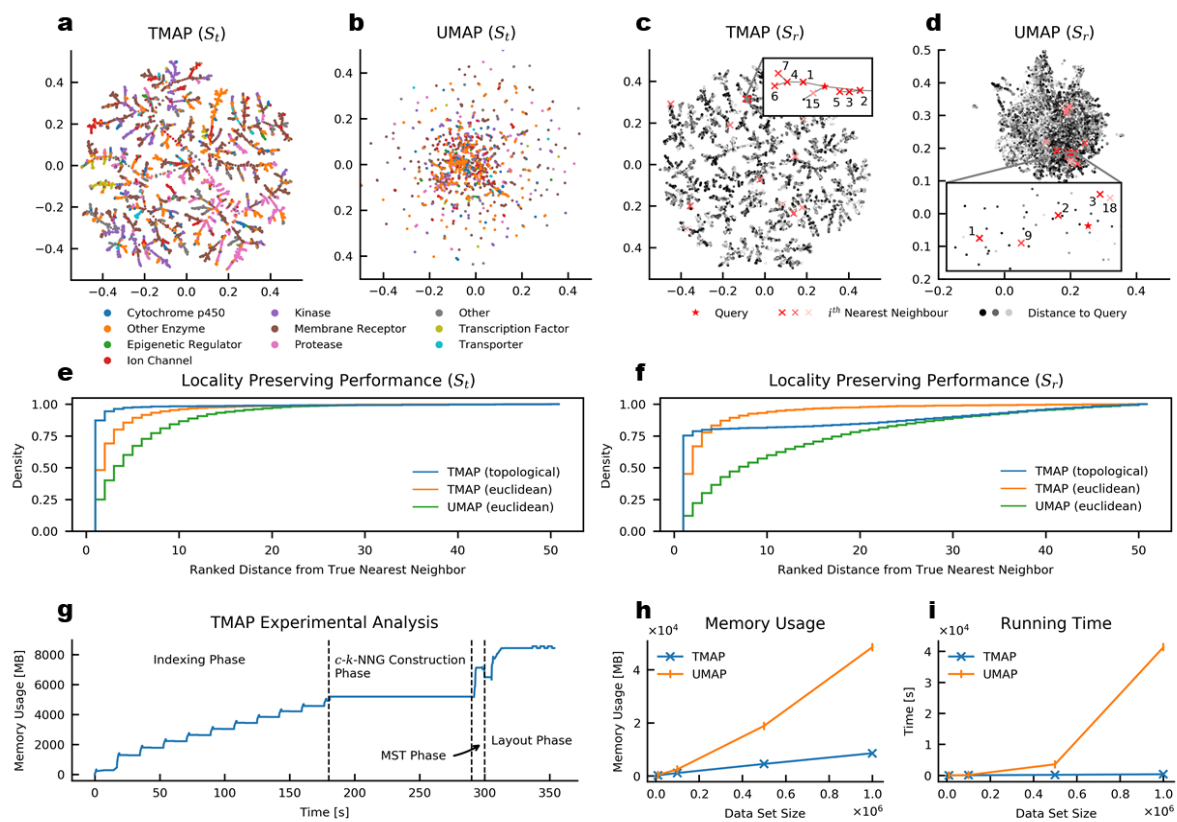

**Fig. 2 Comparing TMAP and UMAP for visualizing ChEMBL**. The first $n$ compounds $S_t$ (**a**, **b**, **e**) and a random sample $S_r$ (**c**, **d**, **f**), each of size $n = 10,000$, were drawn from the 512-D ECFP-encoded ChEMBL data set to visualize the distribution of biological entity classes and k-nearest neighbors respectively. (**a**) TMAP lays out the data as a single connected tree, whereas (**b**) UMAP draws what appears to be a highly disconnected graph, with the connection between components becoming impossible to assert. TMAP keeps the intra- and inter-cluster distances at the same magnitude, increasing the visual resolution of the plot. (**c**, **d**) The 20 nearest neighbors of a randomly selected compound from a random sample. (**c**) TMAP directly connects the query compound to three of the 20 nearest neighbors (1, 2, 15); nearest neighbors 1 through 7 are all within a topological distance of 3 around the query compound. (**d**) The closest nearest neighbors of the same query compound in the UMAP visualization are true nearest neighbors 2, 3, 18, 9, and 1, with 1 being the farthest of the five. (**e**, **f**) Ranked distances from true nearest neighbor in original high dimensional space after embedding based on topological and Euclidean distance for data sets $S_t$ and $S_r$ respectively. (**g**) Computing the coordinates for a random sample ($n = 1,000,000$) highlights the running time behavior of TMAP and allows an inspection of the time and space requirements of the different phases of the algorithm. Four random samples increasing in size ($n = 10,000$, $n = 100,000$, $n = 500,000$, and $n = 1,000,000$) detail the differences in memory usage (**h**) and running time (**i**) between TMAP and UMAP ($t_{TMAP} = 4.865s$, $a_{TMAP} = 0.223GB$; $t_{UMAP} = 20.985s$, $a_{UMAP} = 0.383GB$ and $t_{TMAP} = 33.485s$, $a_{TMAP} = 1.12GB$; $t_{UMAP} = 115.661s$, $a_{UMAP} = 2.488GB$ respectively) ($t_{TMAP} = 175.89s$, $a_{TMAP} = 4.521GB$; $t_{UMAP} = 3,577.768s$, $a_{UMAP} = 18.854GB$ and $t_{TMAP} = 354.682s$, $a_{TMAP} = 8.553GB$; $t_{UMAP} = 41,325.944s$, $a_{UMAP} = 48.507GB$ respectively).

**TMAPs of small molecule data sets: ChEMBL, FDB17, DSSTox, and the Natural Products Atlas**

The high performance and relatively low memory usage of TMAP, as well as the ability to generate highly detailed and interpretable representations of high-dimensional data sets, is illustrated here by interactive visualization of a series of small molecule data sets available in the public domain. In these examples we use MHFP6 (512 MinHash permutations), a molecular fingerprint related to ECFP4 but with better performance for virtual screening tasks and the ability to be directly indexed in an LSH Forest data structure, which considerably speeds up computation for large data sets.[45]

As a first example, we discuss the TMAP of the full data set of the ChEMBL database containing the 1.13 million ChEMBL compounds associated with biological assay data. TMAP completes the calculation within 613 seconds with a peak memory usage of 20.562 GB. Note that approximately half of the main memory usage is accounted for by SMILES, activities, and biological entity classes which are loaded for later use in the visualization. To facilitate data analysis, the coordinates computed by TMAP are exported as an interactive portable HTML file using Faerun, where molecules are displayed using the JavaScript library SmilesDrawer (Fig. 3a).[25,52]

Analyzing the distribution of molecules on the tree shows that TMAP groups molecules according to their structure and their biological activity, accurately reflecting similarities calculated in the high-dimensional MHFP6 space. This is well illustrated for a subset of the map (Fig. 3a, insert). In this area of the map, data points in cyan indicate molecules with a high binding affinity for serotonin, norepinephrine, and dopamine neurotransmitters in two connected branches (right side of inset), while data points in orange show inhibitors of the phenylethanolamine N-methyltransferase (PNMT) (left side of inset), and red and dark blue data points indicate nicotinic acetylcholine receptor (nAChRs) ligands and cytochrome p450s (CYPs) inhibitors, respectively.

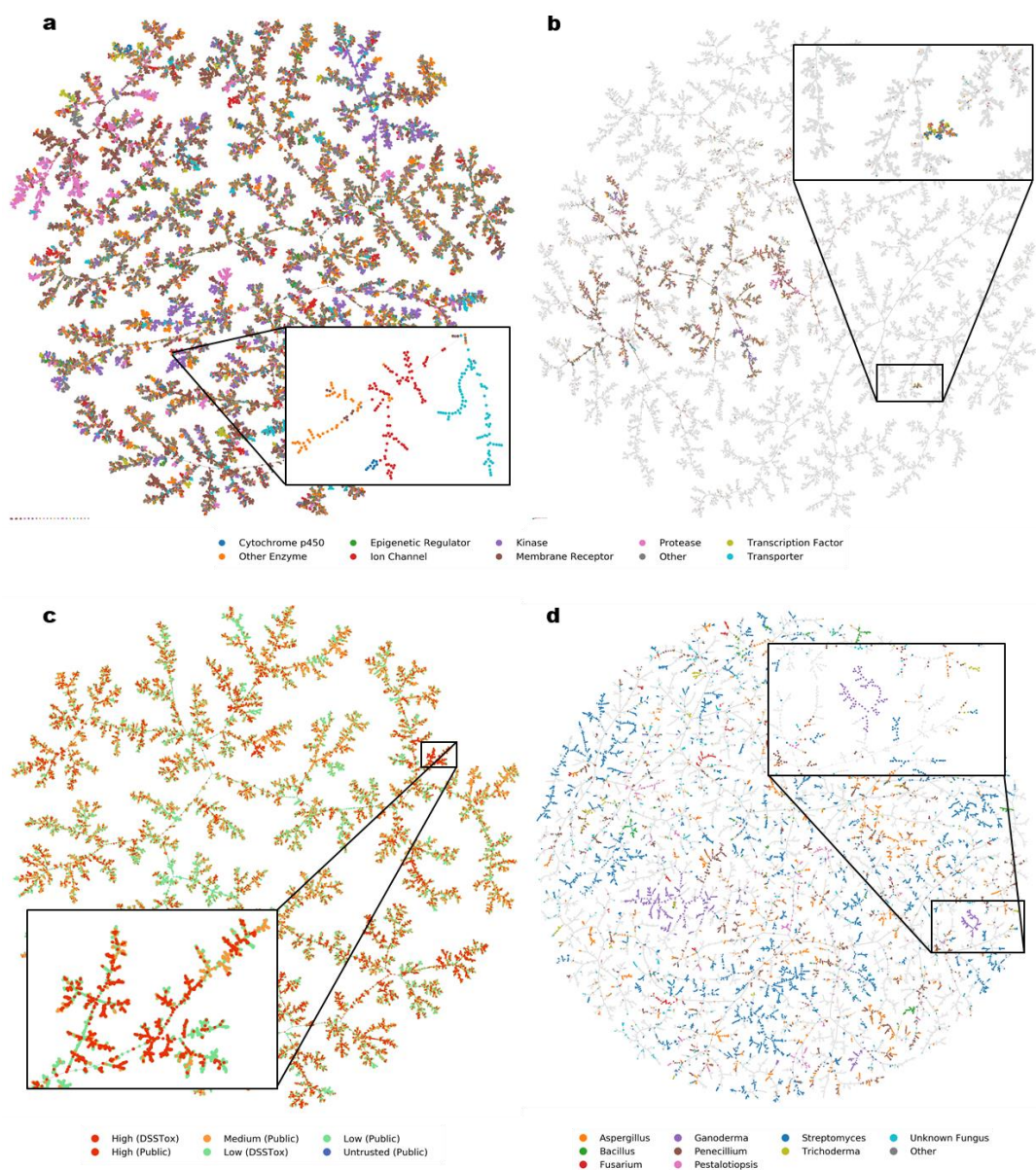

**Fig. 3 TMAP visualization of ChEMBL, FDB17, DSSTox, and the Natural Products Atlas in the MHFP6 chemical space**. Please use the interactive versions at https://tmap.gdb.tools to visualize molecular structures associated with each point. (**a**) Visualization of all ChEMBL compounds associated with biological assay data ($n = 1,159,881$) colored by target class. The inset shows molecules with a high binding affinity for serotonin, norepinephrine, and dopamine neurotransmitters (cyan); inhibitors of the phenylethanolamine N-methyltransferase (orange); and structurally related compounds with high binding affinities for nicotinic acetylcholine receptors and inhibitory effects on cytochrome p450s (red, dark blue). (**b**) The ChEMBL data set was merged with fragment database (FDB17) compounds ($n = 11,261,085$) and visualized. FDB17 molecules are shown in light gray. The inset shows a branch of steroid and steroid-like ChEMBL compounds, as well as dominantly FDB17 branches which are sparsely populated by ChEMBL molecules. (**c**) Visualization of DSSTox compounds colored by reported toxicity level. The inset shows a subtree containing a high number of toxic compounds structurally similar or related to naphthalenes and other polycyclic aromatic hydrocarbons. (**d**) The Natural Products Atlas chemical space colored by origin genus of the 9 largest groups. The inset shows that structurally similar compounds are grouped into distinct branches and subbranches and are usually produced by plants and fungi from the same genus.

As a second example, we visualize the ChEMBL set merged with FDB17 ($n = 10,101,204$) into a superset of size $n = 11,261,085$ (Fig. 3b), which corresponds to the largest data set that TMAP can successfully handle. As above, the TMAP 2D-layout accurately reflects structural and functional similarities computed in the high-dimensional MHFP6 space. In this TMAP visualization, the majority of ChEMBL compounds accumulate in closely connected clusters (branches) due to the prevalence of aromatic carbocycles. A notable exception is a relatively sizable branch of steroids and steroid-like compounds, which is connected to a branch of FDB17 molecules containing non-aromatic 5-membered carbocycles and ketones (Fig. 3b, insert). Many more detailed insights can be gained by inspecting the interactive map in Faerun (http://tmap-fdb.gdb.tools).

Further examples include MHFP6-encoded compounds from the Distributed Structure-Searchable Toxicity (DSSTox) Database ($n = 848,816$) and the Natural Products Atlas ($n = 24,594$). Visualizing DSSTox and coloring the resulting tree by toxicity rating, TMAP creates several subtrees and branches representing structural regions with a high incidence of highly toxic compounds (shown in red, Fig. 3c). An example of such a subtree contains naphthalenes and other polycyclic aromatic hydrocarbons (Fig. 3c, insert). The TMAP tree of the Natural Products Atlas was colored according to origin genus and reveals that branches and subbranches containing distinct substructures usually correlate with a certain genus such as various combinations of phenols, fused cyclopentanes, lactones and steroids produced by the fungi genus Ganoderma (colored purple in Fig. 3d, inset).

**Visualization of the MoleculeNet Benchmark Data Sets**

We further illustrate TMAP to visualize the MoleculeNet, a benchmark for molecular machine learning which has found wide adaption in cheminformatics and encompasses 16 data sets ranging in size and composition (Table 1).[18]  As for the other small molecule data sets above, we computed MHFP6 fingerprints of the associated molecules and the corresponding TMAPs, which we then color-coded according to various numerical values available in the benchmarks. The procedure was applied with all MoleculeNet data sets except for **QM7/b**, where no SMILES have been provided.

The resulting TMAP representations, accessible at the TMAP website (http://tmap.gdb.tools), reveal the detailed structure of the data sets as well as the behaviour of methods applied to these data sets as a function of the chemical structures of the molecules. For example, TMAPs of the **QM8** and **QM9** ($n = 21,786$ and $n = 133,885$), which contain small molecules and DFT-modelled parameters, reveal relationships between molecular structures and the various computed physico-chemical values. For instance the TMAP of the **QM8** data set color-coded by the oscillator strengths of the lowest two singlet electronic states reveals how the value correlates with molecular structure and explains the performance differences in machine learning models trained on Coulomb matrices versus those trained on structure-sensitive molecular fingerprints.[53] In the case of the **ESOL** data set containing measured and calculated water solubility values of common small molecules ($n = 1,128$), its TMAP color-coded with the difference between computed and measured values reveals the limitation of the **ESOL** model when estimating solubility of polycyclic aromatic hydrocarbons and compounds containing pyridines. For the **FreeSolv** data set ($n = 642$) containing small molecules and their measured and calculated hydration free energy in water, the TMAP visualization hints at possible limitations of the method when calculating hydration free energies of sugars. Finally, for the **MUV** data set ($n = 93,087$), which contains active small drug-like molecules against 17 different protein targets mixed in each case with inactive decoy molecules, the various TMAPs reveal differences in the structural distribution of actives among decoys. Actives are usually well distributed but appear to form clusters in certain subsets (e.g. MUV-548 and MUV-846), explaining the generally higher performance of fingerprint benchmarks for these subsets.[47]

**Application to other scientific data sets**

We further illustrate the general applicability of TMAP to visualize data sets from the fields of linguistics, biology, and particle physics. All produced maps are available as interactive Faerun plots on the TMAP website (http://tmap.gdb.tools).

Our first example concerns visualization of the RCSB Protein Data Bank, which contains experimental 3D-structures of proteins and nucleic acids ($n = 131,236$).[54] The PDB files were extracted from the Protein Data Bank and encoded using the protein shape fingerprint 3DP (136-D integer vector, 256 weighted MinHash samples) 3DP encodes the structural shape of large molecules stored as PDB files based on through-space distances of atoms.[22] Processing data extracted from the PDB and indexed using a weighted variant of MinHash, demonstrates the ability of TMAP to visualize both global and local structure, improving on previous efforts on the visualization of the database.[22,55] The global structure of the 3DP-encoded PDB data is dominated by the size (heavy atom count) of the proteins (Fig. 5a), on the other hand, the local structure is defined by properties such as the fraction of negative charges (Fig. 5b).

As an additional example from biology, we consider the PANCAN data set ($n = 800$, $d = 20,531$), which consists of gene expressions of patients having different types of tumors (PRAD, KIRC, LUAD, COAD, and BRCA), randomly extracted from the cancer genome atlas database.[56] Here we index the PANCAN data directly using the LSH Forest data structure and weighted MinHash. The output produced by processing the PANCAN data set displays the successful differentiation of tumor types based on RNA sequencing data by the algorithm (Fig. 5c). We also visualize the ProteomeHD data set using TMAP.[57] This data set consists of co-regulation scores of 5,013 proteins, annotated with their respective cellular location. In addition to the ProteomeHD data set, Kustatscher *et al.* also released an R script to create a map of the set using t-SNE which took a total of 400s to complete; in contrast,

TMAP visualized the data set within 32 seconds (Fig. 5d), successfully clustering proteins by their cellular location based on co-regulation scores. As a further biology example, our TMAP webpage also features flow cytometry measurements ($n = 436{,}877, d = 14$), exemplifying the methods application for the visualization of relatively low dimensional data.[17,58]

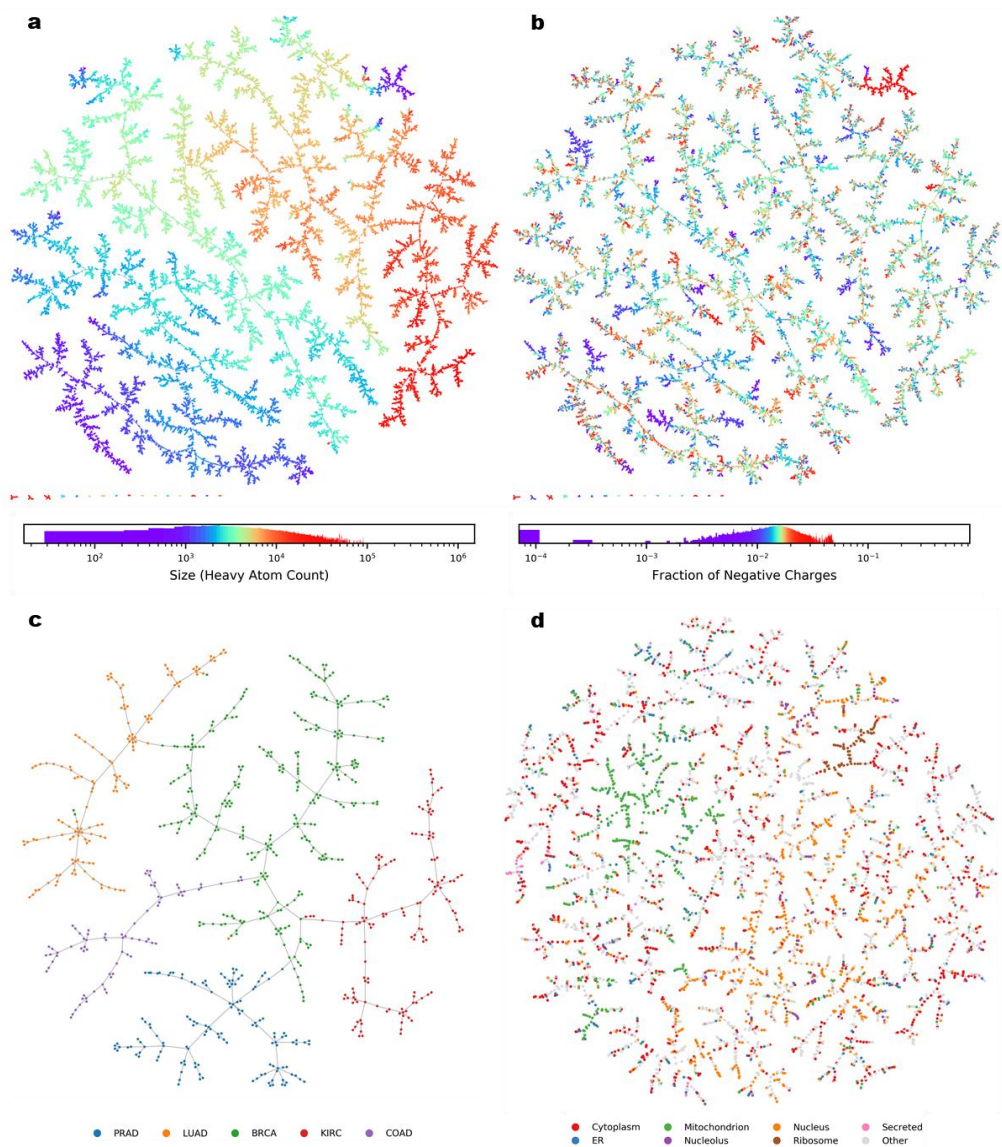

**Fig. 5 TMAP visualizations of the RCSB Protein Data Bank (PDB), PANCAN, and ProteomeHD data**. For (**a**) and (**b**), please use the interactive versions at http://pdb-tmap.gdb.tools to visualize protein structures associated with each point. 3DP-encoded PDB entries visualized using TMAP with weighted MinHash indexing, the color bars show the log-log distribution of the property values. (**a**) Colored according to the macromolecular size (heavy atom count). The resulting map reflects the size-sensitivity of the 3DP fingerprint. (**b**) Colored according to the fraction of negative charges in the molecules. Macromolecules with a high fraction of negatively charged atoms, predominantly nucleic acids, are visible as clusters of red branches. (**c**) The PANCAN data set (n=801, d=20,531) consists of gene expressions data of five types of tumors (PRAD, KIRC, LUAD, COAD, and BRCA) and was indexed using a weighted variant of the MinHash algorithm. (**d**) Visualization of the ProteomeHD data set (n=5,013, d=5,013) based on co-regulation scores of proteins. The data points have been colored according to the associated cellular location.

As an example from physics, we represent the MiniBooNE data set ($n = 130{,}065$, $d = 50$), which consists of measurements extracted from Fermilab's MiniBooNE experiment and contains the detection of signal (electron neutrinos) and background (muon neutrinos) events[59]. As the attributes in MiniBooNE are real numbers, we use the Annoy indexing library which supports the cosine metric in phase I of the algorithm to index the data for $k$-NNG construction, which demonstrates the modularity of TMAP.[60] This example reflects the independence of the MST and layout phases of the algorithm from the input data, displaying the distribution of the signal over the background data (Fig. 6a).

Outside of the natural sciences, we exemplify TMAP to visualize the GUTENBERG set as an example of a data set from linguistics. This data set features a selection of $n = 3{,}036$ books by 142 authors written in English.[61] To analyze this data, we define a book fingerprint as a dense-form binary vector indicating which words from the universe of all words extracted from all books occurred at least once in a given book (yielding a dimensionality of $d = 1{,}217{,}078$), and index this book fingerprint using the LSH Forest data structure with MinHash. The visualization of the GUTENBERG data set exemplifies the ability of TMAP to handle input with extremely high dimensionality ($d = 1{,}217{,}078$) efficiently (Fig. 6b). The works of different authors tend to populate specific branches, with notable expected exceptions such as the autobiography of Charles Darwin, which does not lie on the same branch as all his other works. Meanwhile, the works of Alfred Russel Wallace are found on subbranches of the Darwin branch.

Related to linguistics, the TMAP webpage further features a map of the distribution of different scientific journals (Nature, Cell, Angewandte Chemie, Science, the Journal of the American Chemical Society, and Demography) over the entire PubMed article space ($n = 327{,}628$, $d = 1{,}633{,}762$), perceiving specialization, diversification, and overlaps; as well as a TMAP of the NeurIPS conference papers ($n = 7{,}241$, $d = 225{,}423$), visualizing the increase in occurrence of the word "deep" in conference paper abstracts over time (1987-2016).

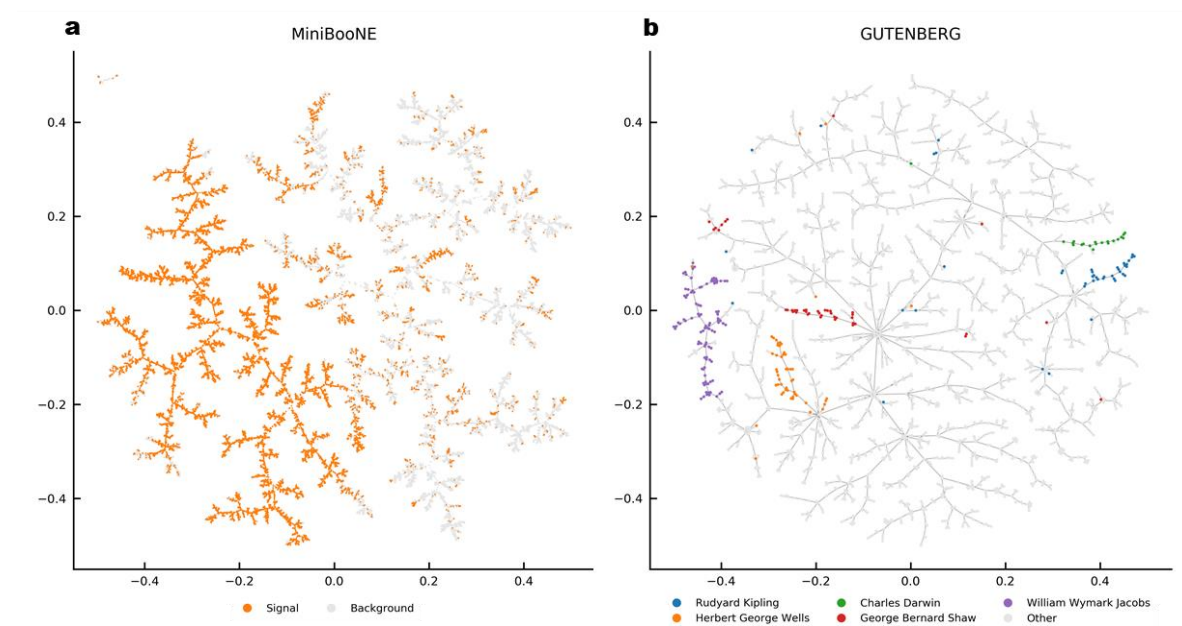

**Fig. 6 Visualizing linguistics, RNA sequencing, and particle physics data sets.** (**a**) The MiniBooNE data set ($n = 130,065$, $d = 50$) consists of measurements extracted from Fermilab's MiniBooNE experiment. TMAP visualizes the distribution of the signal data among the background. (**b**) The GUTENBERG data set is a selection of books by 142 authors ($n = 3,036$, $d = 1,217,078$). The works of five different authors are shown to occupy distinct branches. Interactive version of these maps and further examples can be found at http://tmap.gdb.tools.

## Conclusion

In this study, we introduced TMAP as a visualization method for very large, high-dimensional data sets enabling high data interpretability by preserving and visualizing both global and local features. By using TMAP in combination with the MHFP6 fingerprint, we can visualize databases of millions of organic small molecules and the associated property data with a high degree of resolution, which was not possible with previous methods. TMAP is also well-suited to visualize arbitrary data sets such as images, text, or RNA-seq data, hinting at its usefulness in a wide range of fields including computational linguistics or biology.

TMAP excels with its low memory usage and running time, with performance superior to other visualization algorithms such as t-SNE, UMAP or PCA. By adjusting the available parameters and leveraging output quality and memory usage, TMAP does not require specialized hardware for high-quality visualizations of data sets containing millions of data points. Most importantly, TMAP generates visualizations with an empirical sub-linear time complexity of $O(n^{0.931})$, allowing to visualize much larger high dimensional data sets than previous methods.

All the TMAP visualizations presented, including installation and usage instructions, are available as interactive online versions (http://tmap.gdb.tools). The source code for TMAP is available on GitHub (https://github.com/reymond-group/tmap) and a Python package can be obtained using the conda package manager.

**Acknowledgments.** This work was supported financially by the Swiss National Science Foundation, NCCR TransCure.

**Conflict of interest statement**. The authors declare no conflict of interest.

**Author contributions**. DP designed and realized the study and wrote the paper. JLR supervised the study and wrote the paper.

# Supplementary Information

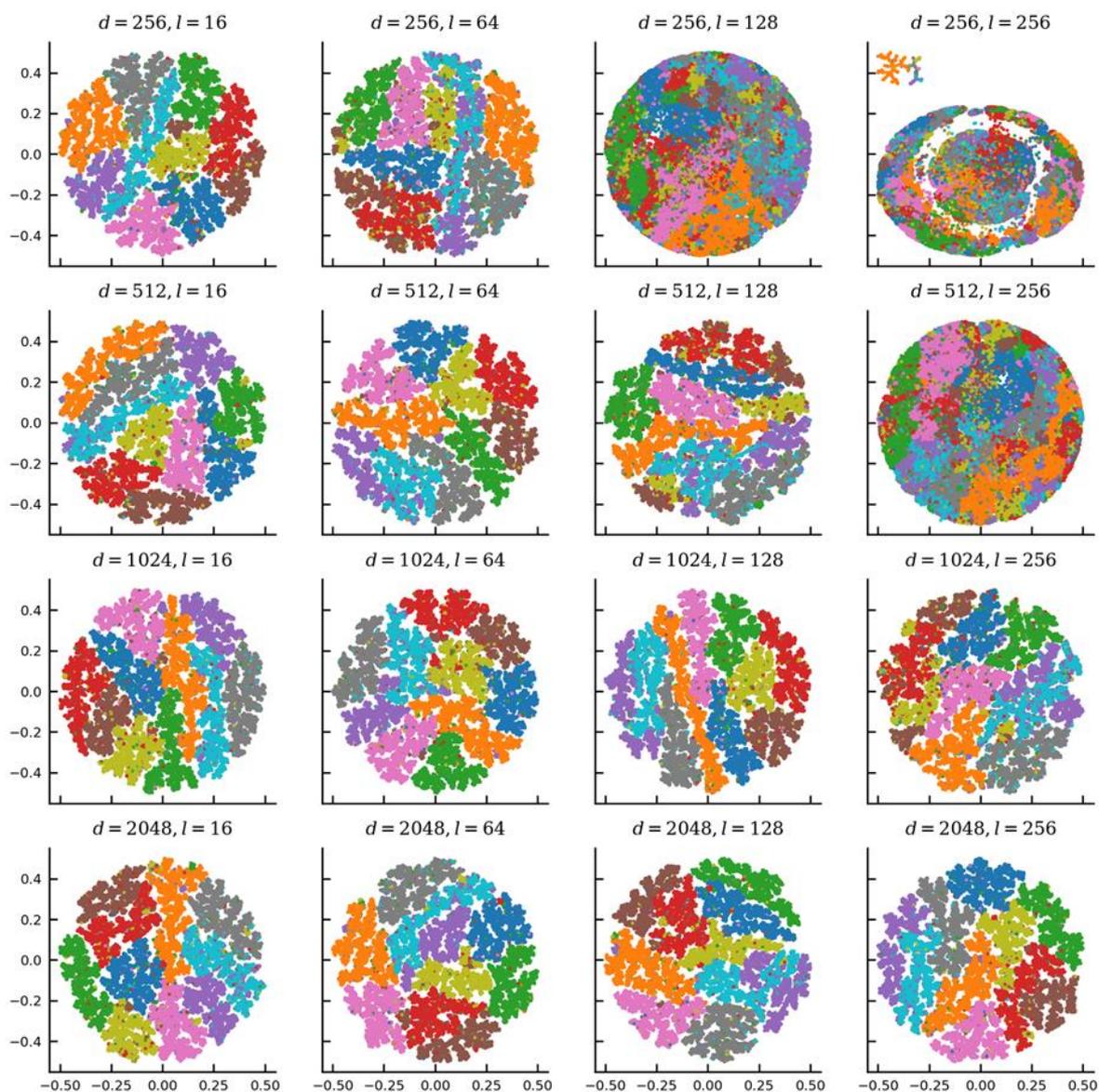

**Fig. S1 Influence of LSH Forest parameters $d$ and $l$ on visualization of MNIST**. While phase I of the algorithm mainly influences the preservation of locality (Fig. S6), extreme values where $d \approx l$ lead to a deterioration of visualization quality.

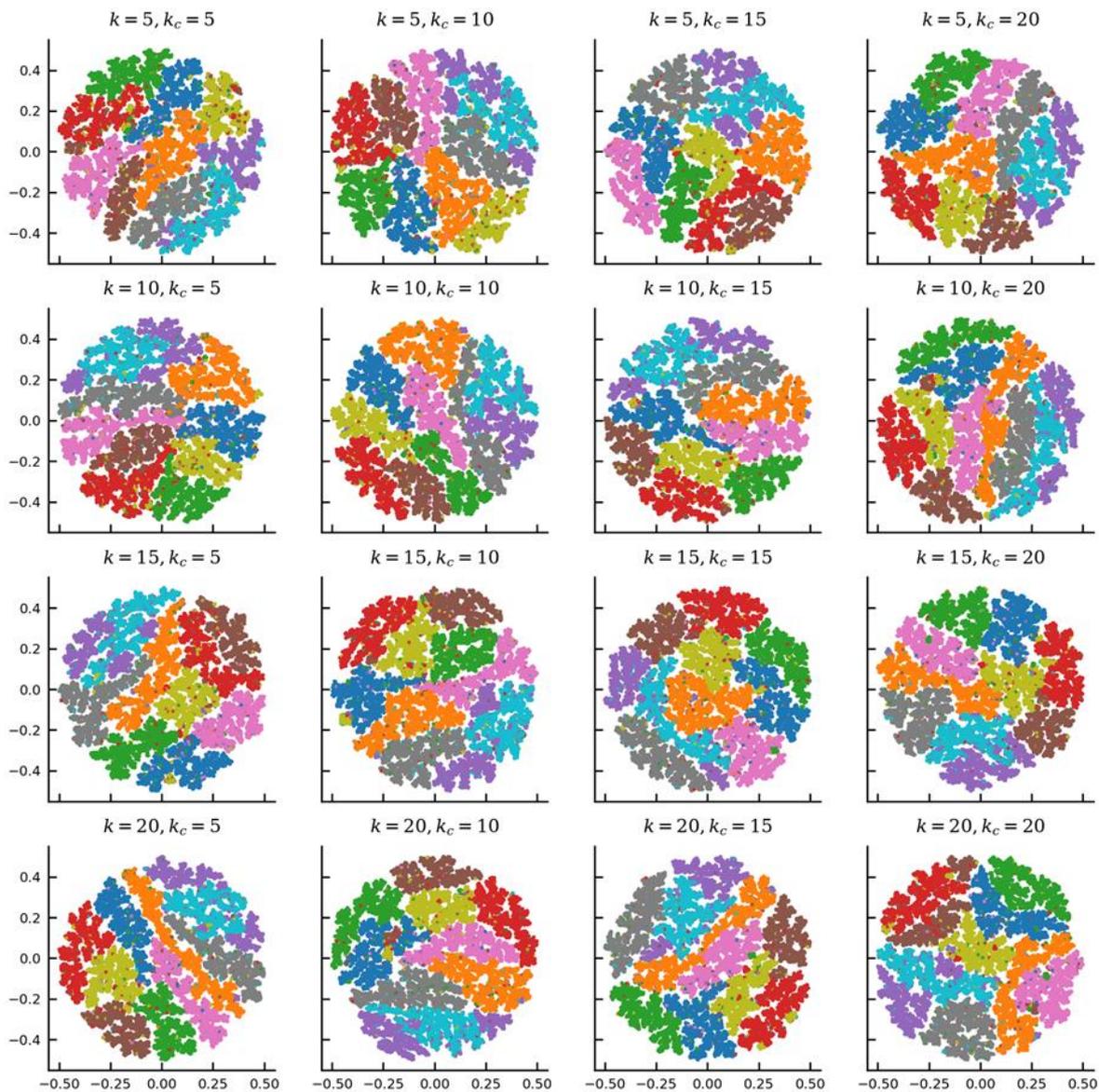

**Fig. S2 Influence of LSH Forest parameters $k$ and $k_c$ on visualization of MNIST**. Whereas parameter $k$ directly influences the average degree of the $k$-nearest neighbor graph, $k_c$ increases the quality of the returned $k$ nearest neighbors. Both parameters only marginally influence the aesthetics and quality of the visualization.

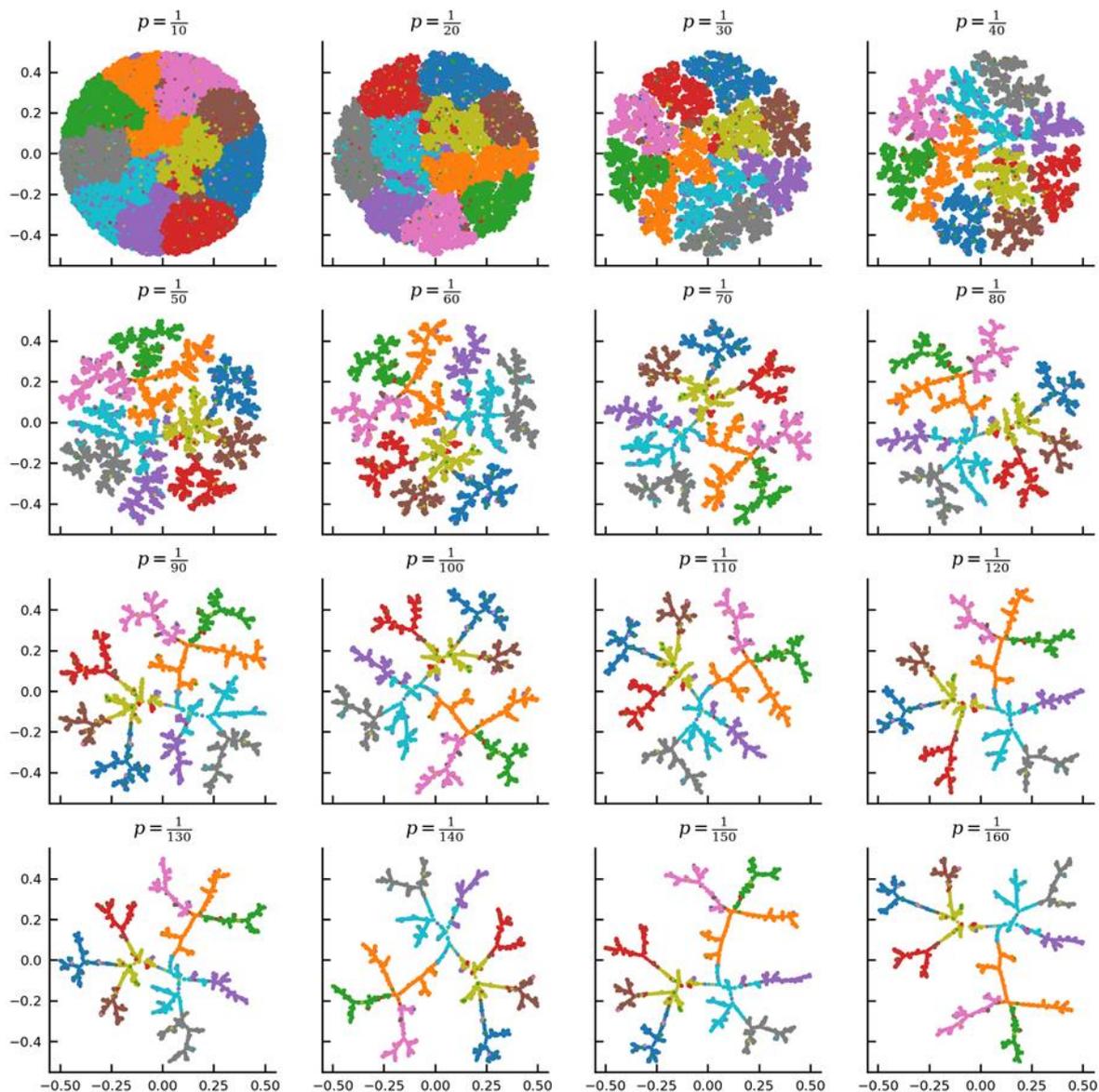

**Fig. S3 Influence of parameter $p$ on visualization of MNIST**. The point size parameter $p$ has major influence on the aesthetics of the visualization, as it controls the sparseness of the drawn tree. Decreasing the point size and thus the repulsive force between two points, allows the layout algorithm to draw points closer to their respective (sub) branches.

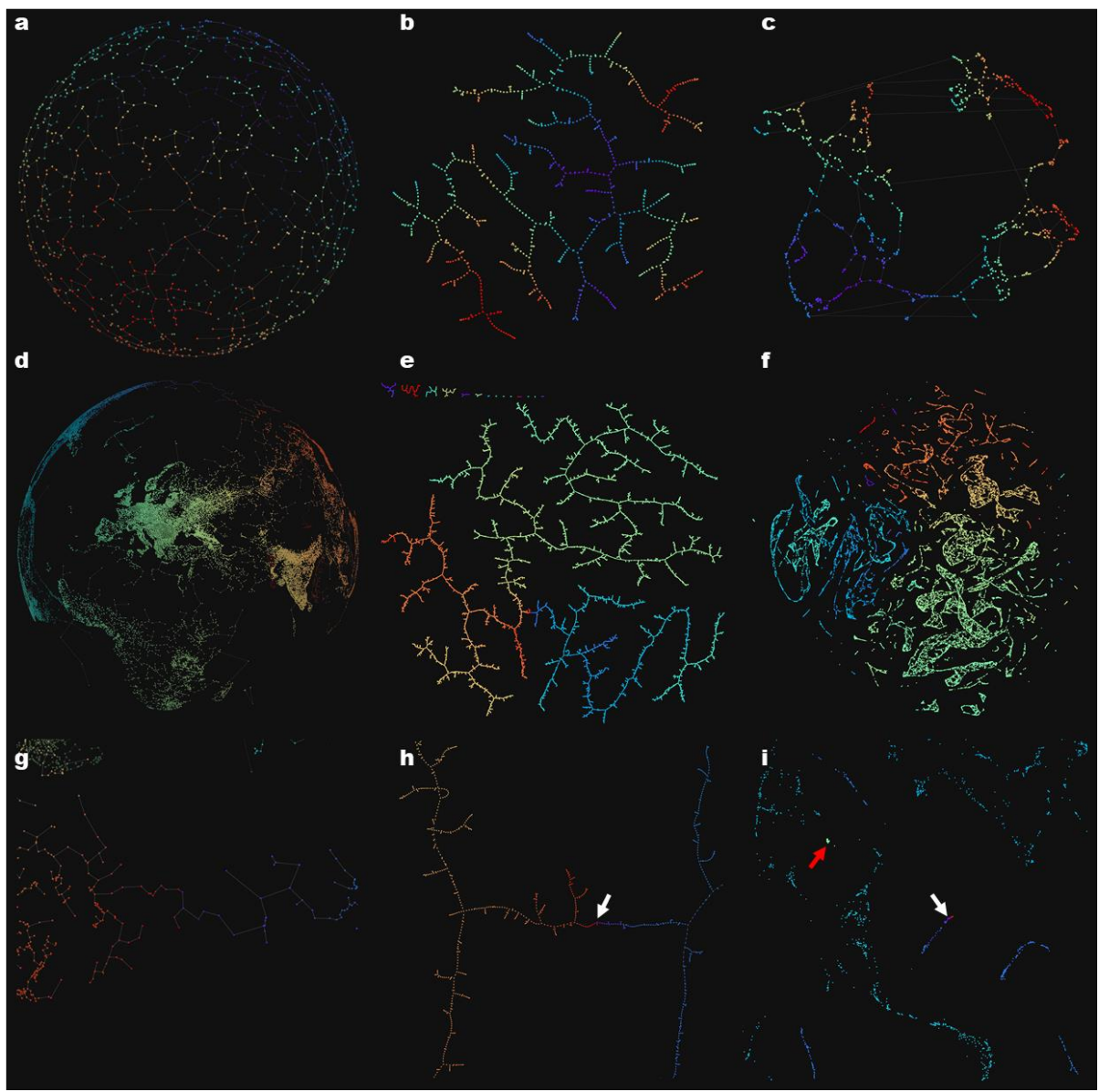

**Fig. S4 Examples of 2D embeddings of points sampled from the surface of 2-spheres by TMAP and UMAP.** (a, b, c) Colored by the points x-components in the original 3-dimensional Euclidean space, (d, e, f) colored by longitude. (a) Randomly distributed points picked from the surface of a 2-sphere were embedded on the two-dimensional plane by TMAP (b) and UMAP (c) using an angular metric in 0.859s and 2.96s wall-clock time respectively. Both algorithms were run with their respective nearest neighbor parameters set to 10. The results clarify the intrinsic need to break locality when embedding unbounded closed surfaces. (d) The location of cities with a population higher than 1,000 ($n = 47980$) mapped to a sphere. (e) TMAP and (f) UMAP embeddings of the city-representing points in a 2D plane. The colors roughly represent Europe and Africa (green), India and central Asia (yellow), east Asia (orange), Oceania (red), and the Americas (blue). Using a Euclidean metric, the execution time of the algorithms was 18.566s and 218.528s, respectively. A detailed view of the Bering Strait (white arrows) in original 3D space with TMAP edges added (g) embedded by TMAP (h) and UMAP (i) highlights the differences between the two approaches. As laying out a graph is vastly more complex than laying out a tree, UMAP produces embedding errors such as the placement of Sardinian cities surrounded by North American cities (red arrow).

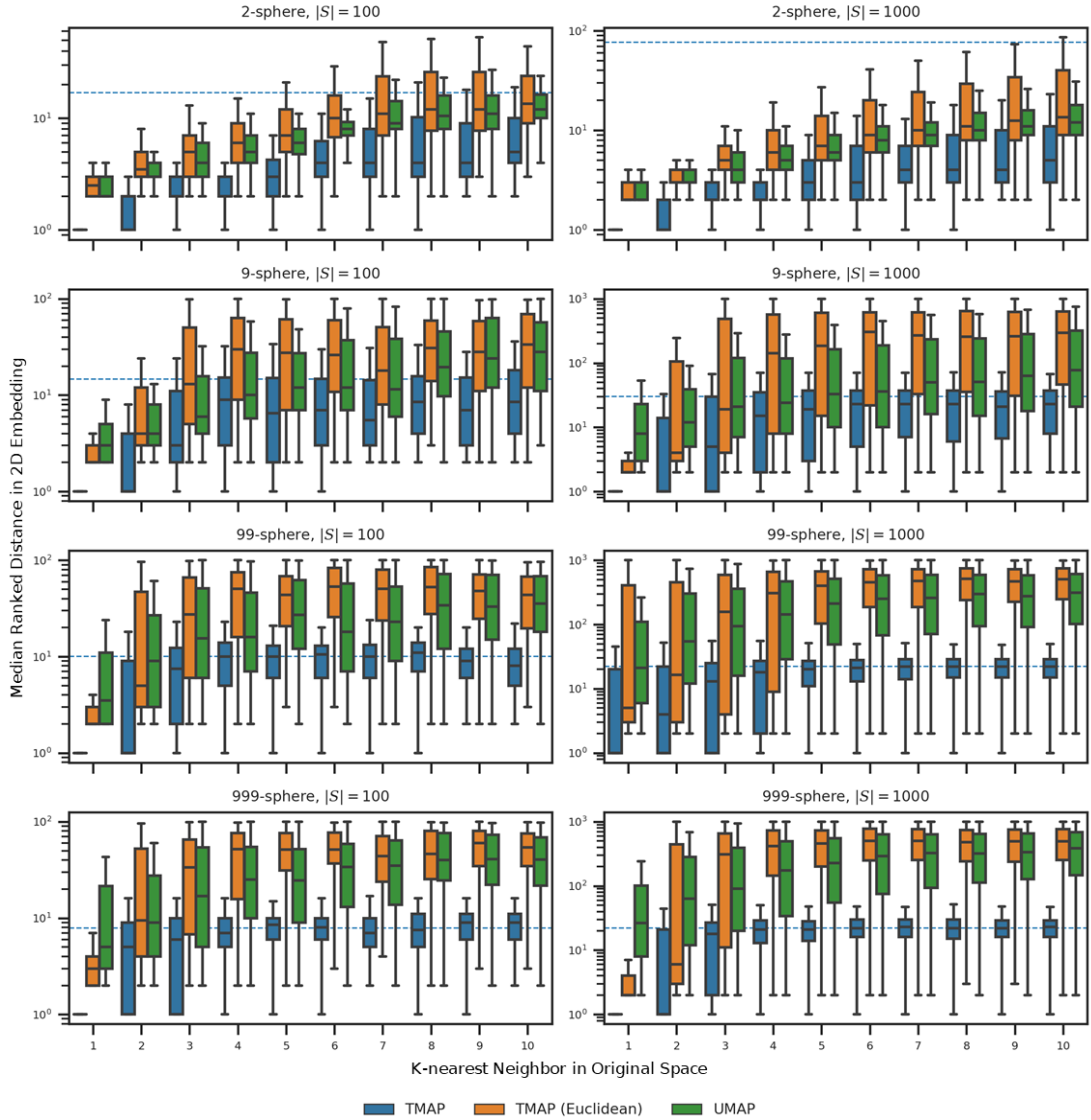

**Fig. S5 Locality-preservation in 2D embeddings of $n$-spheres by TMAP and UMAP**. Random points were uniformly picked from the surface of $(n-1)$-spheres with $n \in \{3, 10, 100, 1000\}$. In addition, the number of points were varied between 100 and 1,000. These examples represent edge cases due to the nature of high dimensional hyperspheres. For each point, the 10 nearest neighbors in the original $n$-dimensional space were compared to their ranked distance in the 2-dimensional plane. As the topological distance cannot be directly compared to the two Euclidean distances given the very high likeliness of collisions at any given (topological) distance due to branching, the average topological distance between any two points was included (blue dashed line).The quality in locality preservation of both TMAP and UMAP degenerates quickly in higher dimensions, however, TMAP tends to preserve the two nearest neighbors even when embedding very high dimensional data.

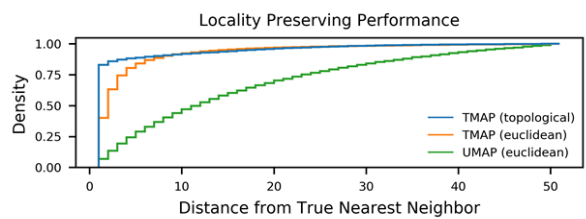

**Fig. S6 Ranked distance from true nearest neighbor when visualizing the MNIST data set**. Ranked distances from true nearest neighbor in original high dimensional space after projection based on topological and Euclidean distance for the MNIST data set. Whereas UMAP preserves less than 10% of true 1-nearest neighbors, TMAP preserves more than 80% based on topological and more than 35% based on Euclidean distance.

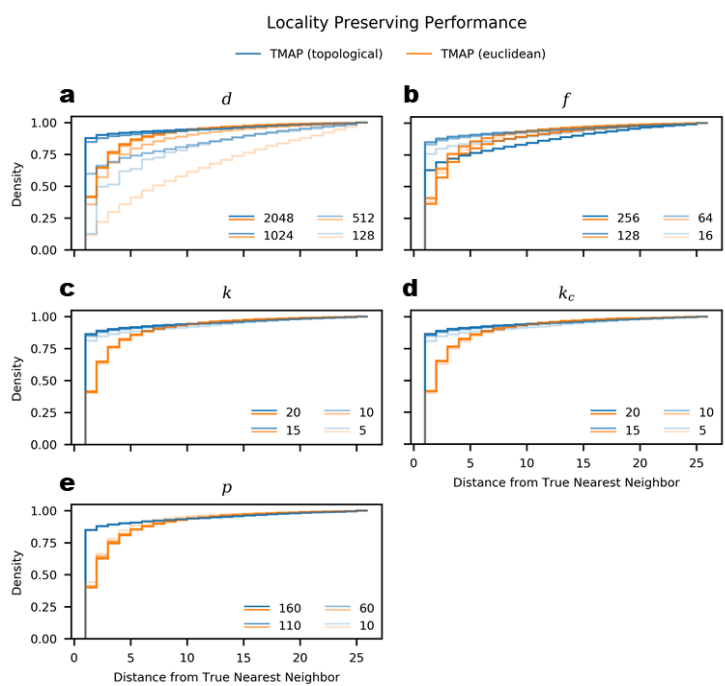

**Fig. S7 Influence of TMAP parameters on locality preserving performance**. Ranked distances from true nearest neighbor in original high dimensional space after projection based on topological and Euclidean distance for the MNIST data set. While, parameters $d$ and $l$ (**a**, **b**) have a major influence on both, the topological and Euclidean measure of locality preserving performance, parameters $k$ and $k_c$ have only marginal influence (**c**, **d**). The point size $p$ does not influence topological distances; however, it has a minor effect on the Euclidean distance-based metric, as higher values increase the sparsity of the drawn tree.

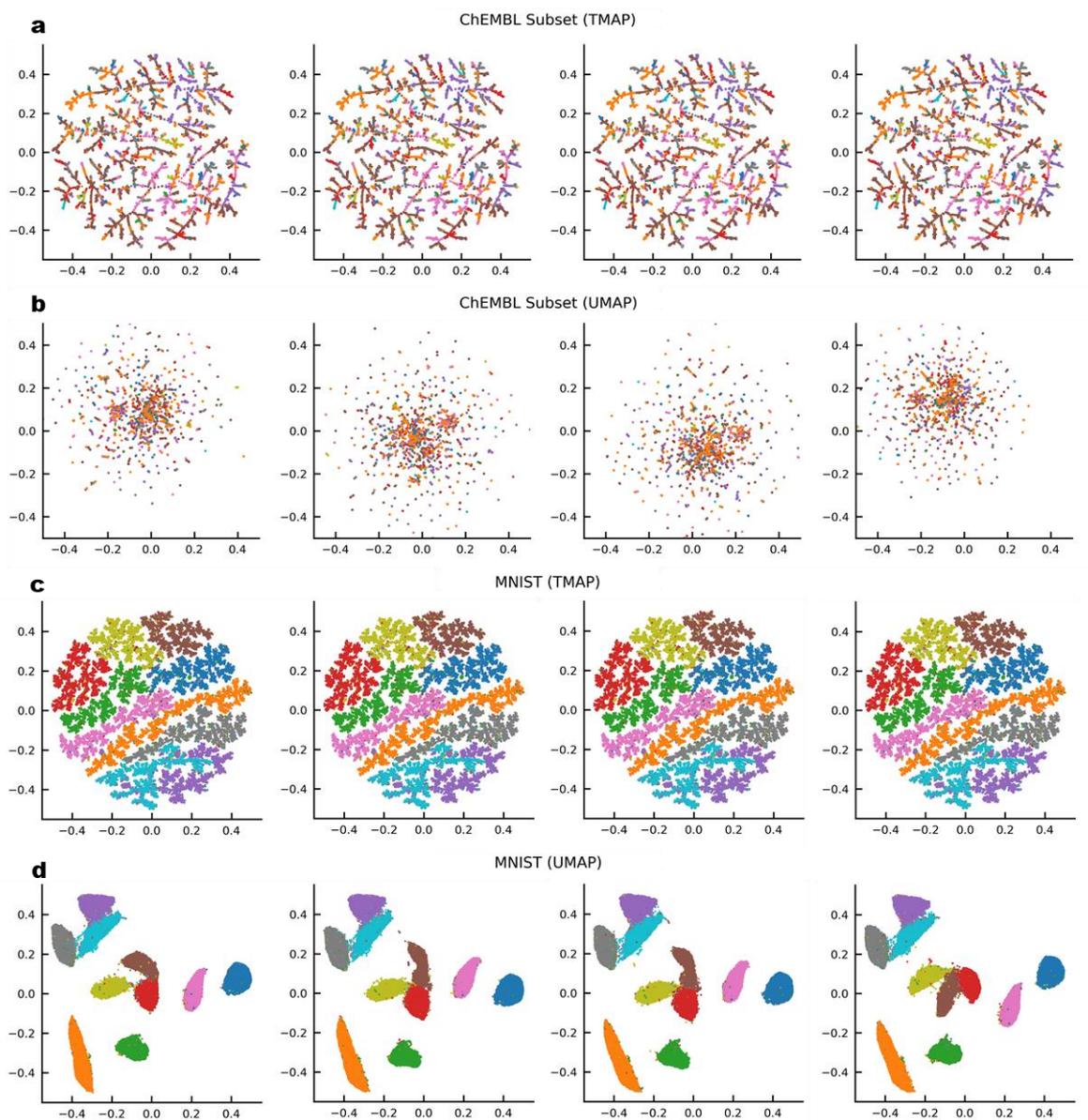

**Fig. S8 Stability of TMAP**. Algorithms TMAP (**a**, **c**) and UMAP (**b**, **d**) have been repeatedly ($n = 4$) run on the same data sets with the same parameters. Whereas the output of TMAP is perceived as identical in all instances, the results yielded by UMAP show considerable differences between each run.